\documentclass[vecphys]{svmult}

\usepackage{makeidx}         
\usepackage{graphicx}        
\usepackage{multicol}        
\usepackage[bottom]{footmisc}

\makeindex             

\begin{document}

\title*{The Connection between Orbits and Isophotal 
Shape in Elliptical Galaxies}
\author{R. Jesseit \inst{1}, 
T. Naab\inst{1} \and A. Burkert\inst{1}}
\institute{Universit\"atssternwarte M\"unchen, Scheinerstr. 1, 
81679 M\"unchen, Germany \texttt{jesseit@usm.lmu.de}}
%
%
\maketitle

We examine the origin of photometrical properties of N-body merger remnants from
the perpective of their orbital content. We show that disc mergers alone are unlikely to form a
boxy-discy dichotomy in isophotal shape, because of the survival of disc-like orbits even in violent mergers.
The shape of the different orbit families can vary strongly, depending on how violent the merger
was. Minor axis tubes can become boxy and box orbits can be round. However, for edge-on projections 
isophotal shape, orbital content and line-of-sight velocity are well connected.

\section{Introduction}
\label{sec:1}
The original classification of elliptical galaxies, which goes back to Edwin Hubble, is based 
on their apparent ellipticity. The projected ellipticity, however, is also a function of inclination, and not
only of the intrinsic ellipticity. \cite{B88} showed that the isophotes of elliptical galaxies 
deviate significantly from perfect ellipses. The deviations can be quantified by the $a_4$-parameter, where
negative $a_4$ values signify boxy deviations and positive $a_4$ values discy deviations from a perfect 
ellipse. But it was also found that ellipticals which have boxy isophotes, are also X-ray bright, more 
luminous, rotate slowly and have cored inner surface density slopes, while discy ellipticals are less luminous, rotate 
fast and have power-law surface density profiles \cite{BDM88}. This led \cite{KB96} to revise the 
Hubble classification of ellipticals using isophotal shape instead of ellipticity as a galaxy family 
membership criterium. 
One of the most successful models of elliptical galaxy formation is the merging of disc galaxies.  
\cite{BA98} proposed that the isophotal shape depends on the violence of the merging and subsequently \cite{N99} 
showed that mergers of galaxies with comparable mass form more likely boxy ellipticals, while mergers
of galaxies with differing mass form discy ellipticals. However, disc-disc mergers do not form a perfect
dichtomy \cite{N03}, while elliptical-elliptical mergers seem to disconnect merging ratio and isophotal 
shape, i.e. they are always boxy \cite{NKB}. In contrast to this gas seems to play an important role in fast-rotating, 
discy ellipticals \cite{N06},\cite{J07}. However, recently the SAURON sample, a survey of 48 
elliptical and lenticular galaxies, \cite{E04} showed that isophotal shapes are not well connected to rotation, i.e. fast rotating 
galaxies can have boxy isophotes \cite{E07}. In the following we want to shed some light  on the connection 
of photometric and kinematic parameters in N-body merger remnants and how they relate to the internal orbital structure.  

\section{Superposition of Orbit Classes and Isophotal Shape}
\label{sec:2}
\begin{figure}
\centering
\includegraphics[height=6cm]{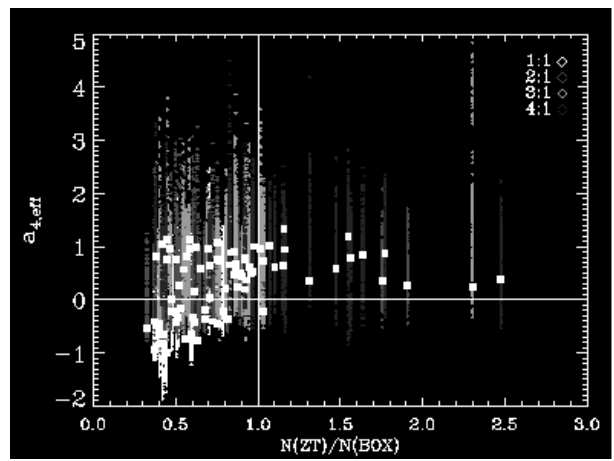}
\caption{Relation between effective isophotal shape and internal orbital structure.White square: most probable
projection of a given remnant. 200 random projections for each remnant.}
\label{fig:1}    
\end{figure}

The orbital structure is the backbone of any galaxy. The properties of different orbit classes 
determines the final shape and kinematic structure of the galaxy. It is much less clear how the
population of orbits is connected to the formation history of a galaxy and if we can detect 
the traces today. We examined a sample of 96 collisioneless disc-disc mergers, with merging mass 
ratios from 1:1 to 4:1. Most of the orbits on which the simulation particles move, belong to one
of the following classes: box orbits, minor axis tubes or major axis tubes. 
In general the abundance of box orbits, which are non-rotating is decreasing with increasing 
merging mass ratio \cite{J05} and vice versa for minor axis tubes. This means that the less violent
the merger is the more particles which have been in the disc of the progenitor survive.
If we correlate the effective isophotal shape with the balance between minor axis tube and box orbits, 
we see that some mergers can appear discy, but have a sizable box orbit component (see Fig. \ref{fig:1}). 

\begin{figure}
\centering
\includegraphics[height=6cm]{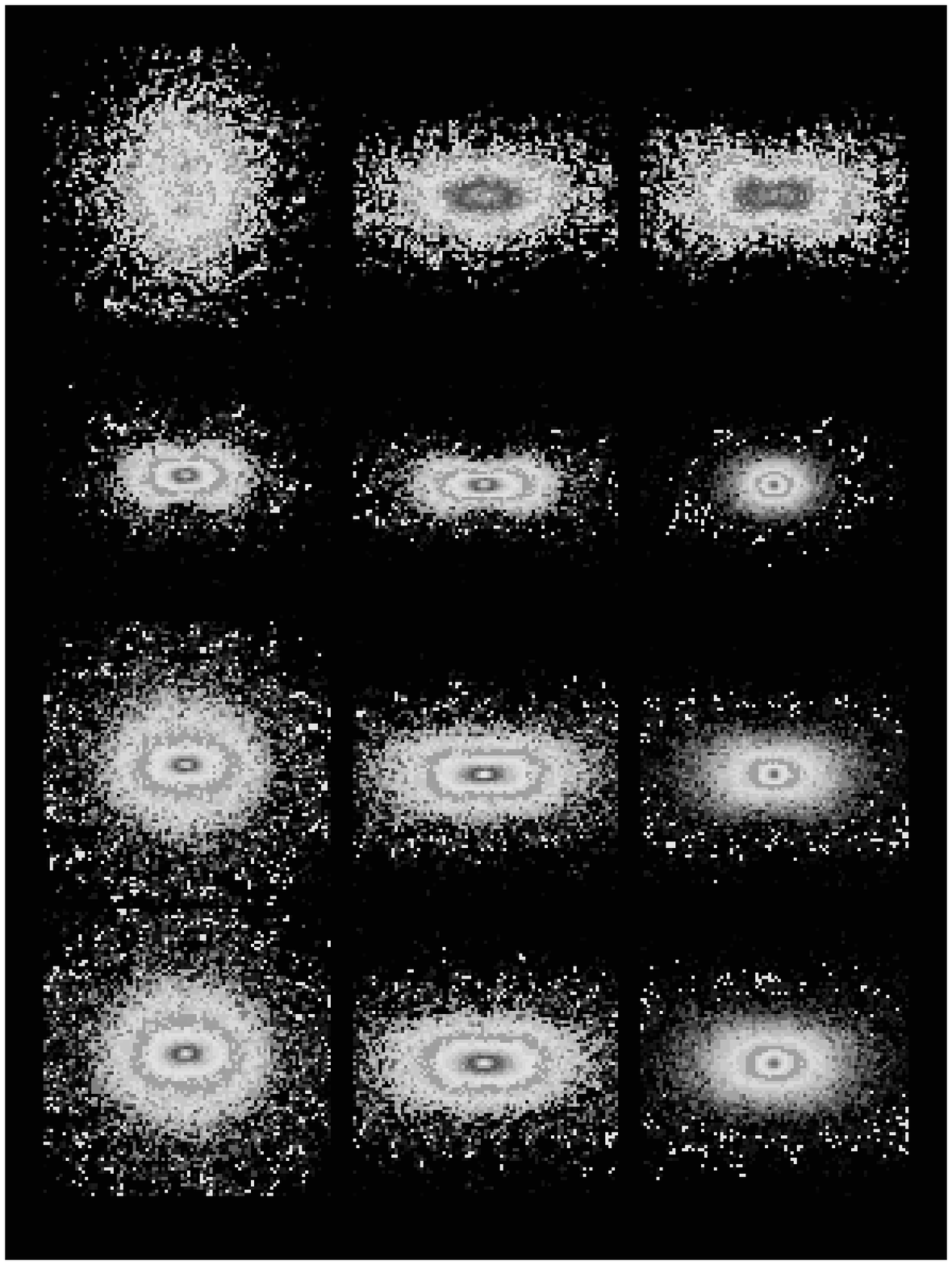}
\includegraphics[height=6cm]{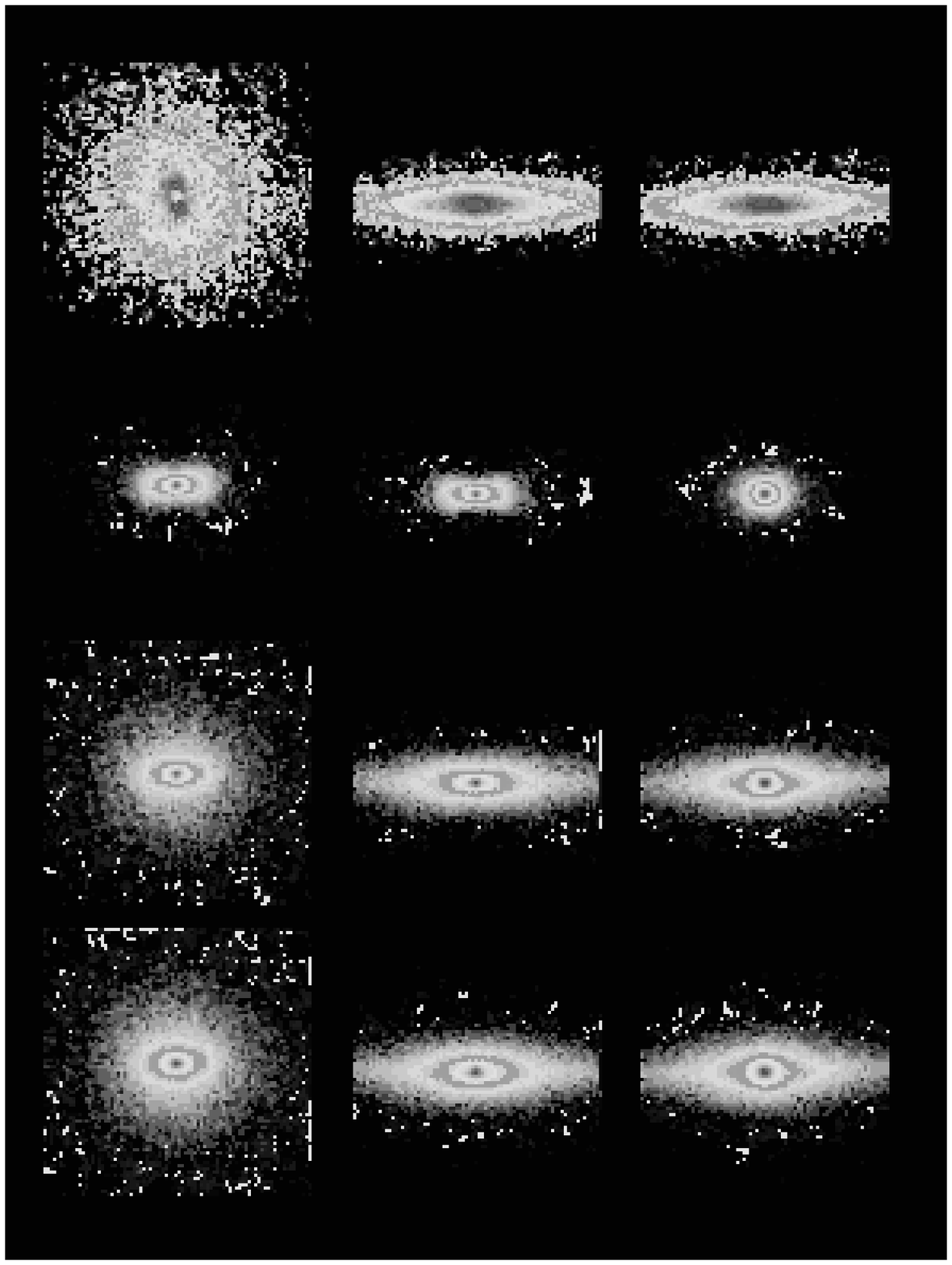}
\caption{2-dimensional shape distribution of particles belonging to a certain 
orbital class. From Top to bottom: Minor axis tubes, box orbits, box orbits + minor axis tubes, all particles. Left panel:
Equal-mass merger remnant with boxy isophotes. Right panel: 4:1 discy merger remnant. Both remnants have approximately
40\%  box orbits.}
\label{fig:2}    
\end{figure}

How can that be? We chose two remnants, one is very boxy and the other is discy, but both have a similar 
fraction of box orbits, i.e. about 40\%. We use the classification to disect the remnants into orbit classes.
Regular orbits or at least orbits who behave regular over very long time scales, respect certain
symmetries even if they move in a complicated triaxial potential, therefore we show the projections along
the principal axes of the inertia tensor (see Fig. \ref{fig:2}). It becomes immediately clear that the spatial
distribution of particles belonging to the same orbit class are very different in each remnant. 
The difference is particularly striking for the minor axis tubes, which can appear peanut-shaped in the 
1:1 remnant and disc-like in the 4:1 remnant. 
 
\begin{figure}
\centering
\includegraphics[height=4cm]{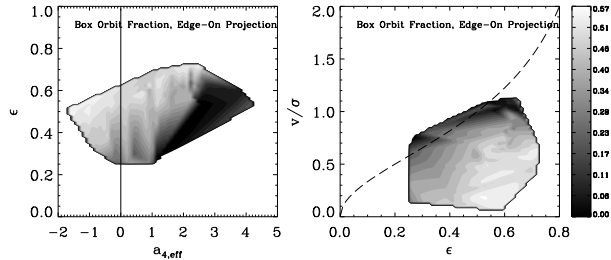}
\caption{Box orbit content of N-body merger remnants in the $a_4$ -$\epsilon$ plane 
and in the $v/\sigma$ -$\epsilon$ plane. The observational parameters are determined
when the merger remnants are seen edge on.  }
\label{fig:3}    
\end{figure}

Although $a_4$  can be an ambiguous indicator for the presence of certain orbit classes, the
kinematic properties, are more clear-cut, e.g. while a box orbit can appear more round or more boxy
depending on the overall shape of the potential the star moves in, the population of stars moving on box orbits
will never show a netto rotation. However, even large rotation can be hidden if we observe 
a galaxy perpendicular to the line-of-sight, i.e. face-on. If we observe our sample of merger remnants 
edge-on we can probably learn the most of their internal structure. In Fig.\ref{fig:3} (left) we see the abundance 
of box orbits with respect to the location of the remnant in the $a_4 -\epsilon$ plane. The isoabundance lines 
cross the $a_4=0$ line, indicating that we can hide a lot of box orbits in a discy remnant. In the $v/\sigma -\epsilon$ 
plane the orbital content can more reliably be deduced. Remnant with little amount of box orbits lie closer
to the isotropic rotator line. The gradient of box orbit content is approximately parallel to this line and 
peaks at $\epsilon=0.5$ and $v/\sigma =0$, where one would expect that triaxial galaxies are located.  

\section{Discussion and Conclusion}
The trends seen here are typical for disc-disc merger remnants, while they are probably representative for low
to intermediate luminosity elliptical galaxies, they cannot explain the whole 
population of elliptical galaxies. The analysis of the orbital fine structure should be extended
to a broader range of formation mechanisms, e.g. E/S0 galaxies with boxy isophotes, will certainly have bars 
\cite{A05} which have not been covered by our present analysis. Dry merging will have certainly played
a role in the formation of the most massive stellar systems in the universe and will lead to round or boxy systems
\cite{NKB},\cite{KB05}, while gas physics is important to form remnants with realistic LOSVDs \cite{N06},\cite{J07}.
Monolithic collapse can also form triaxial self-gravitating systems which can have higher fractions of 
semi-stochastic orbits than our remnants \cite{Aqui07}. Finally binary merger remnants 
can be compared to elliptical galaxies which formed in cosmological simulations and will probably have very different
orbital structures\cite{N07}.

%
%

%
%

\printindex
\end{document}